\documentclass[aps,prl,a4paper,superscriptaddress,twocolumn,showpacs]{revtex4}
\usepackage{graphicx}
\usepackage{latexsym}
\usepackage[]{amsmath}

\begin{document}
\newcommand{\chem}[1]{\ensuremath{\mathrm{#1}}}
\title{Enhancement of Superfluid Stiffness, Suppression of Superconducting $T_{\rm c}$ and Field-induced Magnetism in the Pnictide Superconductor LiFeAs}

\author{F.~L.~Pratt}
\affiliation{ISIS Muon Facility, ISIS, Chilton, Oxon.,
OX11 0QX, United Kingdom}

\author{P.~J.~Baker}
\affiliation{Oxford University Department of Physics, Clarendon Laboratory,
Parks Road, Oxford OX1 3PU, United Kingdom}

\author{S.~J.~Blundell}
\affiliation{Oxford University Department of Physics, Clarendon Laboratory,
Parks Road, Oxford OX1 3PU, United Kingdom}

\author{T.~Lancaster}
\affiliation{Oxford University Department of Physics, Clarendon Laboratory,
Parks Road, Oxford OX1 3PU, United Kingdom}

\author{H.~J.~Lewtas}
\affiliation{Oxford University Department of Physics, Clarendon Laboratory,
Parks Road, Oxford OX1 3PU, United Kingdom}

\author{P.~Adamson}
\affiliation{Inorganic Chemistry Laboratory, University of Oxford,
South Parks Road, Oxford, OX1 3QR, United Kingdom}

\author{M.~J.~Pitcher}
\affiliation{Inorganic Chemistry Laboratory, University of Oxford,
South Parks Road, Oxford, OX1 3QR, United Kingdom}

\author{D.~R.~Parker}
\affiliation{Inorganic Chemistry Laboratory, University of Oxford,
South Parks Road, Oxford, OX1 3QR, United Kingdom}

\author{S.~J.~Clarke}
\affiliation{Inorganic Chemistry Laboratory, University of Oxford,
South Parks Road, Oxford, OX1 3QR, United Kingdom}

\date{\today}

\begin{abstract}
Transverse-field muon-spin rotation measurements performed on two
samples of LiFeAs demonstrate that the superfluid stiffness of the
superconducting condensate in relation to its superconducting transition temperature
is enhanced compared to other pnictide
superconductors.  
Evidence is seen for a field-induced magnetic state in a sample with a significantly suppressed superconducting transition temperature. 
The results in this system highlight the role of direct Fe-Fe
interactions in frustrating pairing mediated by antiferromagnetic-fluctuations and suggest that, in common with other pnictide superconductors, the system is close to a magnetic instability.
\end{abstract}


\maketitle 


The relationship between critical temperature $T_{\rm c}$ and
superfluid stiffness $\rho_{\rm s}$ in a superconductor (SC) provides
important information concerning the balance between the strength of
the pairing and the efficiency of electromagnetic screening.  In many
SCs these two parameters are related by Uemura
scaling~\cite{uemura89,uemura91}, 
though deviations from this are
found for overdoped cuprates \cite{bernhard} and organic
SCs \cite{flpsjb}.  The recently discovered oxypnictide
SCs~\cite{kamihara08} containing FeAs layers show a wide
range of $T_{\rm c}$
\cite{chen08nature,chen08prl,ren08epl,rotter08,sasmal08arxiv} and
those studied so far have shown behavior broadly consistent with Uemura
scaling~\cite{luetkens08prl,drew08prl,carlo08arxiv,luetkens08arxiv,khasanov08arxiv,takeshita08arxiv,drew08arxiv,aczel08arxiv,goko08arxiv},
as observed for hole-doped cuprates.  In this Letter we test this
relationship for LiFeAs \cite{wang08arxiv,pitcher08arxiv,tapp08prb}, a
newly discovered variant of pnictide SC without the
lanthanide-oxide layer.  In LiFeAs the Fe--Fe separation in the FeAs
layers are significantly shorter than in previously studied
pnictide SC compounds.  This produces additional frustrating magnetic
interactions which may weaken the strength of the antiferromagnetic
coupling through Fe--As bonds which has been suggested to mediate the
superconducting pairing \cite{kuroki08}.  We find that this produces a
departure from the previously observed scaling behavior which can provide some insight into the
nature of the pairing in this family of compounds.

LiFeAs crystallizes in a tetragonal space group (P4/nmm) with
$a=0.378$\,nm and $c=0.635$\,nm and contains FeAs layers, based on
edge-sharing tetrahedral FeAs$_4$ units, interspersed with layers
containing Li ions.  The tetrahedra are compressed in the basal plane
relative to those in LaFeAsO and SrFeAs$_2$.  This implies that although the
Fe--As bond distance 0.2414\,nm in LiFeAs is similar to the other
compounds, the Fe--Fe distance of 0.268\,nm is considerably shorter.
Superconductivity can be obtained at up to 18\,K in this compound, 
though differences have been observed even between samples with similar cell volumes, 
probably connected with slight compositional variation, as described previously for compositions close to LiFeAs \cite{Juza68}.
In contrast with the oxypnictides, no
further doping is necessary to induce superconductivity and
the spin-density wave (SDW) state appears to be notably absent from this system.

Transverse field muon spin rotation (TF-$\mu$SR) is a method of accurately 
measuring the internal magnetic field distribution within the vortex lattice (VL)
of a type-II SC~\cite{sonier00}. Spin polarized positive muons
(gyromagnetic ratio $\gamma_{\mu}/2\pi = 135.5$~MHzT$^{-1}$, lifetime $\tau_{\mu} = 
2.2~\mu$s) are implanted into the bulk of the sample, at random positions on the 
length scale of the VL. A magnetic field $B_{c1} < B_0 < B_{c2}$ 
is applied perpendicular to the initial muon spin direction and the muons 
precess around the total local magnetic field at their stopping site. This 
gives a method of randomly sampling the magnetic field distribution using the 
time evolution of the muon spin polarization $P_{x}(t)$, which is related to the 
distribution of local magnetic fields in the sample $p(B)$, via:
\begin{equation}
P_{x}(t) = \int^{\infty}_0 p(B)\cos(\gamma_{\mu}Bt+\phi){\rm d}B,
\label{polarization}
\end{equation}
where $\phi$ is a phase offset associated with the emitted positron
detector geometry.  The function $p(B)$ allows the extraction of the
in-plane penetration depth $\lambda_{ab}$ and hence the superfluid
stiffness $\rho_{\rm s}\propto \lambda_{ab}^{-2}$.

Powder samples of LiFeAs were prepared 
from high purity elemental reagents ( $>$ 99.9 \%) by the methods described in 
references \onlinecite{pitcher08arxiv} and \onlinecite{tapp08prb} and the presence of superconductivity was confirmed by SQUID magnetometry. 
Two samples were selected for the $\mu$SR studies. 
Sample~1 with $T_{\rm c} = 16$~K  and lattice parameters $a$=3.774(1)~\AA, $c$=6.353(1)~\AA~and V=90.5(1)~\AA$^3$ was prepared by heating a 1:1:1 ratio of the elements in a sealed tantalum tube at 750~$^{\circ}$C for 24 hours, a method similar to that described by Tapp et. al. \cite{tapp08prb}. Sample~2 with $T_{\rm c} \sim 12$~K and a broader superconducting transition than that observed for other reported samples was the same as Sample~2 in Ref. \onlinecite{pitcher08arxiv} ($a$=3.774(1)~\AA, $c$=6.354(2)~\AA~and V=90.5(1)~\AA$^3$). 
Laboratory powder X-ray diffraction indicated that the samples were at least 98~\% phase pure.
Measurements using $\mu$SR are not generally sensitive to low level impurity phases and we note also that common impurity phases found in pnictide SCs, such as FeAs and FeAs$_{2}$,
give very different $\mu$SR signals \cite{baker08arxiv} from those which we will describe
below.  
Both samples are stoichiometric 
and have equal cell parameters within experimental uncertainty; we attribute the differences in $T_{\rm c}$ to very small differences in Li occupation which are not readily detectable using structural or bulk analytical probes.
Muon-spin rotation measurements were
made on the GPS instrument at the Swiss Muon Source, Paul Scherrer
Institut, CH.

\begin{figure}[tb]
\includegraphics[width=8.3cm]{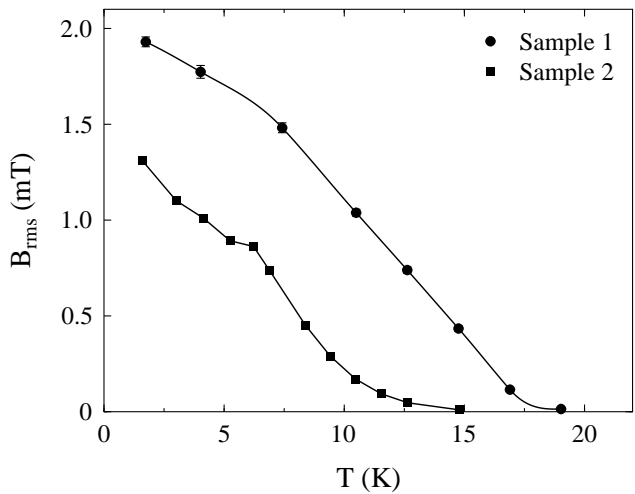}
\caption{The temperature dependence of the superconducting
contribution to the field width of the $\mu$SR lineshape measured
in a transverse field of 40 mT for the two samples of LiFeAs. The curves are a guide to the eye.}
\label{fig1}
\end{figure}

\begin{figure}[ht]
\includegraphics[width=8.3cm]{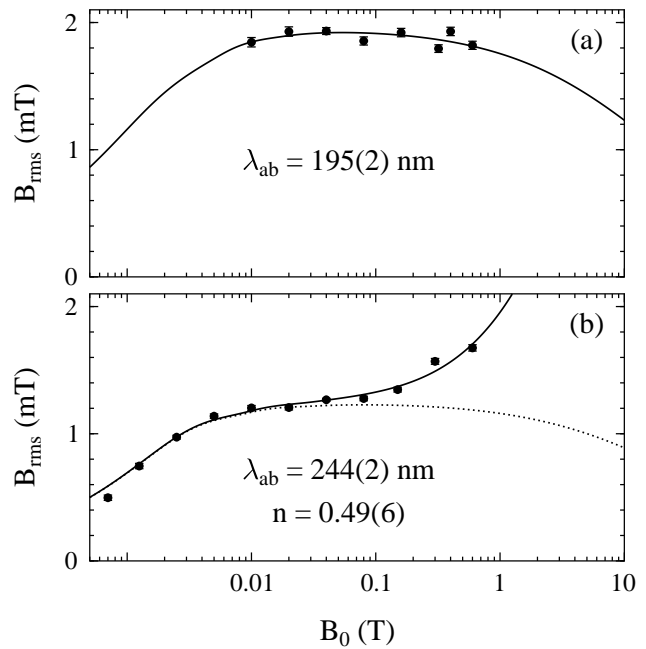}
\caption{(a) The magnetic field dependence of the superconducting
contribution to the $\mu$SR line width for sample 1 measured at 1.6 K.
The field dependence is very weak and the $ab$ penetration depth can
easily be extracted from the data by fitting to the field dependence
of the GL model (solid curve).  (b) The field dependence for sample 2
(with lower $T_{\rm c}$).  The low field behaviour can again be fitted
to the GL model but a marked deviation sets in at fields above 0.1 T
and the overall fit (solid line) needs to include a field-induced
magnetic contribution $\sigma_{\rm M} \propto B^n$ in addition to the
VL contribution (dotted line).  }
\label{fig2}
\end{figure}

In TF-$\mu$SR measurements performed above $T_{\rm c}$, one might
expect $p(B)=\delta(B-B_0)$.  However, broadening of $p(B)$ is
caused by the contribution from randomly oriented nuclear moments near
the muon stopping sites.  This leads to a roughly Gaussian relaxation
$P_x(t) \propto \exp \left[ - \left( \sigma_{\mathrm{n}} t
\right)^{2} / 2 \right]\cos(\gamma_\mu B_0 t + \phi)$.  Below
$T_{\rm c}$ the spectrum $p(B)$ broadens considerably due to the
dephasing contribution from the VL. 
The additional broadening is fitted well by a further Gaussian damping term 
so that the overall damping is 
\begin{equation}
\sigma^2 = \sigma_{\rm VL}^2 + \sigma_{n}^2
\label{eqn0}
\end{equation}
and the corresponding field width due to the VL is $B_{\rm rms} =
\sigma_{\rm VL} / \gamma_\mu$.  The data fitting procedure also takes
into account the small fraction of the signal coming from muons
stopping in the silver sample support which gives a weakly relaxing
background component.  The resulting temperature dependent
contribution to the field width $B_{\rm rms}$ due to the
superconducting VL is shown in Fig.~1 for the two samples in an
applied transverse field of $B_0=40$\,mT.  In both samples, $B_{\rm
rms}$ increases monotonically on cooling below the $T_{\rm c}$.  The
increasing width does not appear to be saturating at low temperatures
and consequently cannot be described by a simple two-fluid or s-wave
BCS model.

The $ab$ plane penetration depth $\lambda_{ab}$ and corresponding superfluid stiffness can be derived from $B_{\rm rms}$ for a powder sample using the relation \cite{Fesenko91}
\begin{equation}
B_{\rm rms} = \frac {\sigma_{\rm VL}} {\gamma_\mu} = (0.00371)^{1/2} \frac {\phi_0} {(3^{1/4} \lambda_{ab})^2}
\label{eqn1}
\end{equation}
Equation \ref{eqn1} is valid in the London limit at fields well above
$B_{c1}$ and well below $B_{c2}$ where $B_{\rm rms}$ is independent of
applied field. A more sophisticated approach allows for the field
dependence that occurs for realistic sample parameters and we use a
recent calculation of the field dependent $B_{\rm rms}$ in the
Ginzburg-Landau (GL) model for an anisotropic SC \cite{brandt03} to
fit our field dependent data (Fig.2).  The original calculation is for
normal field configuration and an extension to the polycrystalline
case was made under the assumption that the length scales $\lambda$
and $\xi$ diverge following $1/\cos \theta$ as the field orientation
approaches the plane at $\theta=0$ (high anisotropy limit), with the
corresponding contribution to the overall width scaling as $\cos
\theta$.  The data obtained for sample 1 is seen to be only very
weakly field dependent (Fig.2a), indicating that the measured field
range is well separated from $B_{c1}$ and $B_{c2}$ and a best fit
\cite{powdernote} yields $\lambda_{ab}=195(2)$\,nm.  In contrast,
sample 2 behaves rather differently [Fig.2(b)].  At lower fields it
follows the GL dependence for the expected longer penetration depth,
however above $\sim$0.1 T a significant increase in the width with
applied field is apparent, suggesting a field-induced magnetic
contribution which can be fitted by adding an additional term
$\sigma_{\rm M}^2$ to Eq.~(\ref{eqn0}), where $\sigma_{\rm M}$ follows
a $B_0^n$ power law. After
allowing for this field induced magnetism, the $\lambda_{ab}$ value in
this second sample is obtained to be 244(2) nm.  

\begin{figure}[t]
\includegraphics[width=8.3cm]{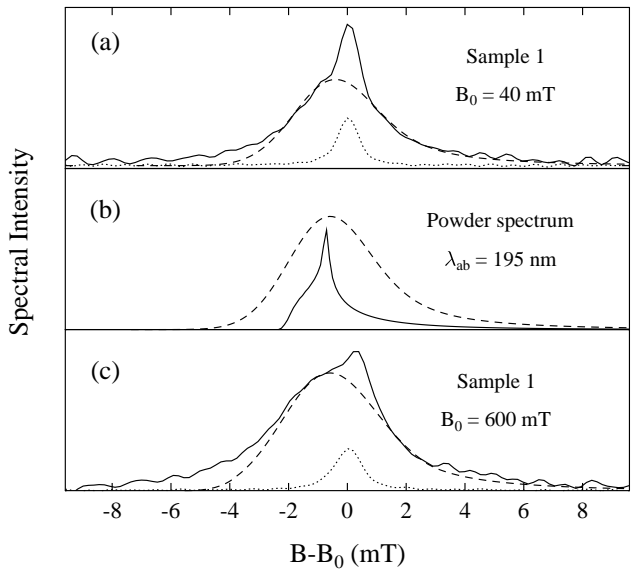}
\caption{(a) Spectral distribution of the internal field in sample 1 at 40 mT obtained by Fourier transformation. The dotted line is the normal state and the solid line is the SC state. The dashed line represents a broadened powder profile detailed in (b),
where the solid line is an appropriate powder profile broadened by a Gaussian to give the dashed curve.
Panel (c) is as for (a) but for a field of 600 mT.}
\label{fig3_s1}
\end{figure}

\begin{figure}[t]
\includegraphics[width=8.3cm]{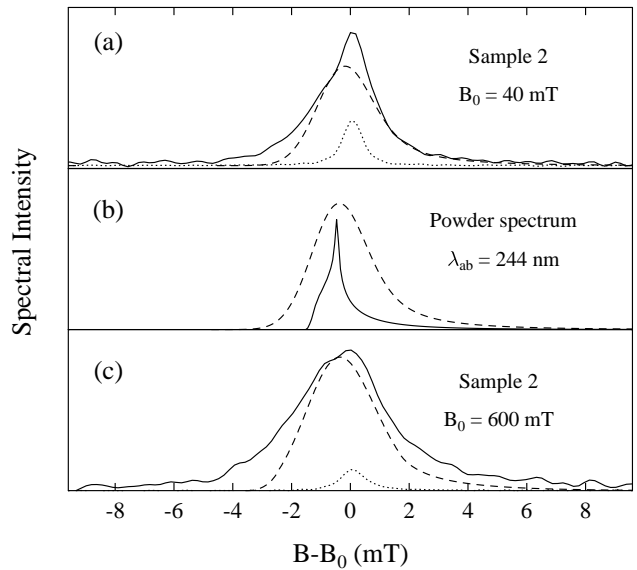}
\caption{As Fig.\ref{fig3_s1} but for sample 2 which shows evidence of field-induced magnetism. 
The VL linewidth is narrower here than for sample 1 and the additional field-induced broadening is clearly seen on both sides of the main peak when comparing the 600 mT data in (c) with the 40mT data in (a). }
\label{fig3_s2}
\end{figure}

Fig.~\ref{fig3_s1}(a) shows the SC field distribution in sample 1 at 40 mT obtained from a Fourier transform.
A broadened VL powder distribution (Fig.~\ref{fig3_s1}(b)) along with a narrow background peak can account for most of the profile, although there is some additional spectral weight on the low field side. 
The profile remains similar when the field is raised to 600 mT (Fig.~\ref{fig3_s1}(c)).
The same spectral analysis for sample~2 is shown in Fig.~\ref{fig3_s2}. 
The profile at 40 mT is qualitatively similar to that seen in sample 1, but with the additional weight on the low field side being relatively larger.
At 600\,mT, however, significant additional broadening becomes apparent on both sides of the peak.

The observation of the departure from GL behavior [Fig.~2(b)] and
shifts in spectral weight [Fig.~\ref{fig3_s2}(c)] for sample 2 at high field may be suggestive
of field-induced magnetism.  
The superconducting state in pnictides is
known to be proximate to SDW order, but no SDW state has been observed directly in the LiFeAs system.  
Field induced magnetism has been investigated in
cuprate SCs, particularly following neutron studies of
La$_{2-x}$Sr$_{x}$CuO$_4$ (LSCO) \cite{lake02nature}, the results of
which have been interpreted as microscopic phase coexistence of SDW
and SC states, driven by coupling of the two order parameters
\cite{demler01prl}.  Such an approach predicts an induced SDW moment
which scales roughly as $B_0^{1/2}$ for $B_0\ll B_{c2}$
\cite{demler01prl}, in agreement with our fitted $n=0.49(6)$.  That the
field-induced broadening shows up in sample 2 and not sample 1 may be
due to the greater proximity to the SDW state driven by some small
change in Li-site occupancy.  This is suggestive that the SC state in
LiFeAs may be extremely close to a magnetic instability.

\begin{figure}[htb]
\includegraphics[width=8.3cm]{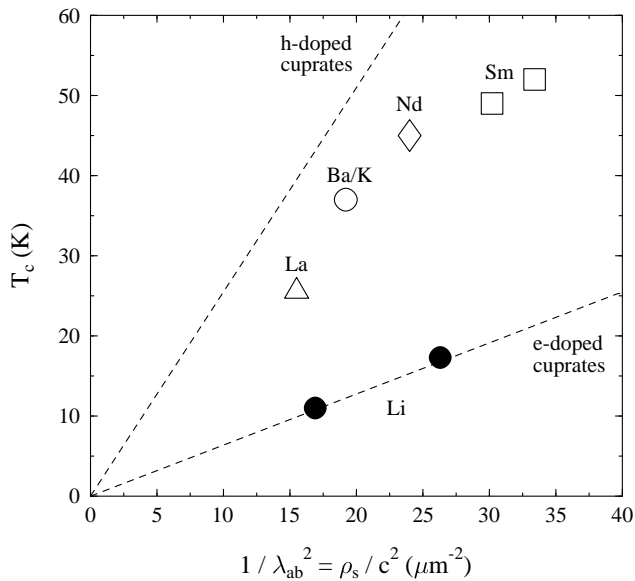}
\caption{An Uemura plot of the superconducting transition temperature $T_{\rm c}$ versus the low temperature superfluid stiffness.
Data obtained here for the two LiFeAs samples are compared with previously reported
data for doped members of the LnFeAsO$_{1-x}$F$_x$ family: Ln =
La\cite{luetkens08prl}, Nd\cite{carlo08arxiv} and Sm
\cite{drew08prl,khasanov08arxiv} and the Ba$_x$K$_{1-x}$Fe$_2$As$_2$
family \cite{aczel08arxiv,goko08arxiv}. The data taken from the previous reports has been restricted to the highest $T_{\rm c}$ sample in each study. 
The lower dashed line indicates the trend line for electron-doped cuprates, which closely corresponds to the behaviour of the LiFeAs system obtained here, in contrast to the other pnictide superconductors, which sit closer to the trend for h-doped cuprates. The symbol size represents the typical estimated uncertainty of the data points.}
\label{fig4}
\end{figure}

The extracted values of $\lambda_{ab}$ allow us to place these
materials on an Uemura plot, as shown in Fig.~\ref{fig4}, alongside
values obtained from reported data on other FeAs-based SCs using
Eq.~(\ref{eqn1}).  In a region where standard Uemura scaling applies
there will be a linear relation between $\rho_s$ and $T_{\rm
c}$. However, from Fig.~\ref{fig4} it can be seen that no single
scaling line can be used for all of the data.  The data appear to
split into two groups with the trend for the LiFeAs samples being lower
than any trend line that would describe the LaFeAsO$_x$F$_y$ and Ba$_x$K$_{1-x}$Fe$_2$As$_2$
data.  This indicates that the LiFeAs samples have a superfluid
stiffness that is enhanced over the values that might be expected on
the basis of their observed transition temperatures and the behaviour of
the other FeAs SCs.  Alternatively one may regard $T_{\rm c}$ for a
given strength of superfluid condensate as being suppressed in the
LiFeAs case, particularly when compared to the magnetic lanthanides
NdFeAsO$_{1-x}$F$_x$ and SmFeAsO$_{1-x}$F$_x$.  This implies that
although the superconducting state in the FeAs layers is reasonably
robust (the superfluid is stiff) the strength of the pairing is
significantly suppressed for LiFeAs.  If the pairing is mediated by
antiferromagnetic fluctuations, originating from the Fe--As--Fe exchange
interactions, we may speculate that this mechanism may be frustrated
by the increasing role played by direct Fe-Fe interactions in LiFeAs.
The Fe--As--Fe exchange is also expected to be modified by the compression of the
FeAs$_4$ tetrahedra and consequent changes to the Fe--As--Fe bond angles in the LiFeAs structure. 
Associated changes in the electronic structure will also increases the electronic bandwidth \cite{singh08} and hence contribute to a
reduction in the effective pairing strength. The increased bandwidth
also leads to a reduced effective mass $m^*$, so that the superfluid
stiffness $n_s/m^*$ is then expected to become enhanced relative to $T_{\rm c}$, as we
have observed experimentally in this study.

Part of this work was performed at the S$\mu$S Swiss Muon Source, Paul Scherrer Institute,
Villigen, CH. We are grateful to Hubertus Luetkens for experimental assistance and
to the EPSRC (UK) for financial support.


\begin{thebibliography}{xx}
\bibitem{powdernote}
Note that the quoted errors in $\lambda_{ab}$ are statistical and do
not include the additional systematic error due to approximating the
powder lineshape by a Gaussian,  however these errors are common to
all powder measurements and we keep to the simpler powder
approximation to allow relative comparisons to be made for powder data
across the pnictide family.
\end{thebibliography}
\end{document}